\documentclass[a4paper,pra,aps,superscriptaddress,citeautoscript,twocolumn,twoside,showpacs,longbibliography]{revtex4-1}

\usepackage{amsmath}
\usepackage{graphicx}
\usepackage{amssymb}
\usepackage{epsfig}
\usepackage{amsfonts}
\usepackage{color}

\renewcommand{\S}{\mathrm{S}}

\newcommand{\eg}{{{e.g.}}}
\newcommand{\ie}{{{i.e.}}}
\newcommand{\etal}{{\it{et al.}}}
\newcommand{\US}{\mathrm{US}}

\newcommand{\ket}[1]{| #1 \rangle}
\newcommand{\bra}[1]{\langle #1 |}
\newcommand{\oprod}[1]{\ket{#1}\bra{#1}}

\newcommand{\beq}{\begin{eqnarray}}
\newcommand{\eeq}{\end{eqnarray}}

 \def\>{\rangle}

\newcommand{\Tr}{\operatorname{tr}}%from Miguel

\begin{document}

\title{Temporal steering and security of quantum
key distribution \\ with mutually-unbiased bases against
individual attacks}

\author{Karol Bartkiewicz}
\email{bark@amu.edu.pl} \affiliation{Faculty of Physics, Adam
Mickiewicz University, PL-61-614 Pozna\'n, Poland}
\affiliation{RCPTM, Joint Laboratory of Optics of Palack\'y
University and Institute of Physics of Academy of Sciences of the
Czech Republic, 17. listopadu 12, 772 07 Olomouc, Czech Republic }

\author{Anton\'{i}n \v{C}ernoch}
\affiliation{Institute of Physics of Academy of Science of the
Czech Republic, Joint Laboratory of Optics of Palack\'y University
and Institute of Physics of Academy of Sciences of the Czech
Republic, 17. listopadu 50A, 77207 Olomouc, Czech Republic}

\author{Karel Lemr}
\affiliation{RCPTM, Joint Laboratory of Optics of Palack\'y
University and Institute of Physics of Academy of Sciences of the
Czech Republic, 17. listopadu 12, 772 07 Olomouc, Czech Republic }

\author{Adam Miranowicz}
\affiliation{CEMS, RIKEN, 351-0198 Wako-shi, Japan}
\affiliation{Faculty of Physics, Adam Mickiewicz University,
PL-61-614 Pozna\'n, Poland}

\author{Franco Nori}
\affiliation{CEMS, RIKEN, 351-0198 Wako-shi, Japan}
\affiliation{Department of Physics, The University of Michigan,
Ann Arbor, MI 48109-1040, USA}

\begin{abstract}
Temporal steering, which is a temporal analogue of
Einstein-Podolsky-Rosen steering, refers to temporal quantum
correlations between the initial and final state of a quantum
system. Our analysis of temporal steering inequalities in relation
to the average quantum bit error rates reveals the interplay
between temporal steering and quantum cloning, which guarantees
the security of quantum key-distribution based on
mutually-unbiased bases against individual attacks. The key
distributions analyzed here include the Bennett-Brassard 1984
protocol (BB84) and the six-state 1998 protocol by Bruss (B98).
Moreover, we define a temporal steerable weight, which enables us
to identify a kind of monogamy of temporal correlations that is
essential to quantum cryptography and useful for analyzing various
scenarios of quantum causality.
\end{abstract}

% 03.67.Mn Entanglement measures, witnesses, and other characterizations
% 42.50.Dv Quantum state engineering and measurements in quantum optics
% 89.70.+c Information theory and communication theory

\date{\today}
\maketitle

%------------------------------------------------------------------
\section{Introduction}

Quantum steering, also known as Einstein-Podolsky-Rosen (EPR)
steering, refers to quantum correlations, which enable one system
to nonlocally steer (or affect) another system by using only local
measurements. This concept was introduced by
Schr\"odinger~\cite{Schrodinger35} over 80 years ago as a
generalization of the EPR paradox~\cite{EPR35} and quantum
entanglement. Surprisingly, our understanding and applications of
this phenomenon are still very limited despite of breathtaking
research and progress (see, e.g., Ref.~\cite{EPR80}). For example,
a few recent experimental demonstrations of EPR steering were
reported in Refs.~\cite{Saunders10, Walborn11,Wittmann12,
Smith12,Bennet12, Handchen12, Steinlechner13, Su13,Schneeloch13},
including even a loophole-free experiment~\cite{Wittmann12}. This
research has also been devoted to analyzing the relations and
potential applications of EPR steering to secure quantum
communication (see, e.g., recent
Refs.~\cite{He15,Pusey15,Kocsis15} and references therein).

Temporal steering (TS) is a temporal analog of EPR steering, and
introduced only very recently~\cite{Chen14}. Our understanding of
this phenomenon is even more
limited~\cite{Chen14,Li15,Karthik15,Mal15,Chen15,Chen16,Chiu16}.

Here we study TS in the context of quantum cryptography and
fundamental issues of quantum theory including quantum cloning and
Heraclitus ``panta rhei''~\cite{Plato}. Here we study the relation
between TS and two quantum key distribution (QKD) protocols: BB84
by Bennett and Brassard~\cite{BB84} and the six-state 1998
protocol by Bruss (B98)~\cite{Bruss98}. These two QKD protocols
constitute a class of so-called mutually unbiased bases (MUB)
protocols for qubits. By studying these examples in detail, we
explain the reason for the security of the MUB protocols based on
temporal steering against individual attacks. Our results show
that the unconditional security of these
protocols~\cite{secBB84,secB98,GisinRMP} implies the existence of
a kind of monogamous temporal correlations, which we refer to as
temporal steering monogamy. If the protocols are not
unconditionally secure, it would not be possible to distinguish
the following two cases: whether the same original particle or two
different particles (\eg, the original and its copy) are observed
at two different moments in time. This would mean that the famous
phrase of Heraclitus ``No man ever steps in the same river
twice''~\cite{Plato}, taken literally, could be fundamentally true
as there would be no way of checking if there exists a single
reality of particles observed at two moments in time. We note that
some relations of TS and BB84 have already been found in the
original paper on TS by Chen \etal~\cite{Chen14}. Our analysis
reveals a much deeper relation of TS and secure communication.

In this paper, we also propose and apply a measure of TS, which
can be referred to as a TS weight in analogy to the EPR steering
weight introduced in Ref.~\cite{Skrzypczyk14}. We note that TS and
its various measures can also have other non-cryptographic
applications in, \eg, quantifying strong non-Markovianity, as
recently shown in Ref.~\cite{Chen16}.

This paper is organized as follows: In Sec.~II, we present a
theoretical framework for analyzing TS. In Sec.~III, we derive TS
inequalities for bit-error rates. In Sec.~IV, we introduce the TS
weight. We conclude in Sec.~V.

%------------------------------------------------------------------
\section{Theoretical framework}

To set up the theoretical framework for TS, let us start with the
assumption that Alice and Bob share a secret sequence of qubit
(spin-$\frac{1}{2}$) observables that they will measure.
Alternatively, they can select their observables at random and
reject those not matching. Alice prepares her states from an
unknown torrent of qubits by performing a Stern-Gerlach-type
experiment~\cite{Gerlach22} with photons using a polariser.
Specifically, Alice separates qubits of opposite values of the
analysed spin observable. Let us assume that Alice and Bob use MUB
(see, \eg, Ref.~\cite{Wootters89}) and their observables are the
Pauli matrices $A_{1}=B_{1}=\sigma _{1}$, $A_{2}=B_{2}=\sigma
_{2}$, and $A_{3}=B_{3}=\sigma _{3}$. The observables $A$ and $B$
have two eigenvalues $a=\pm1$ and $b=\pm1$, respectively. In our
experiment we identify the observables $A_i=B_i$ with the
corresponding Pauli operators $\sigma_i$. Alice chooses her
observable $A_{i}=\sigma_i$ and its value $a=\pm 1$ by a proper
polarisation rotation of the polarised photon. This state
preparation strategy is equivalent to Alice performing her
measurement at time $t_A=0$ on a photon described by an
appropriate polarisation statistics, and then sending the detected
eigenstates of the observable to Bob. The qubit that is delivered
to Bob is a conditional state that depends on Alice's choice of
the observable, her outcome, and the initial state of the
transmitted two-level system. The initial state is imposed by
Alice's choice of measurement and her results. The final measured
state is an outcome of the state evolution and the specific
measurement setting used by Bob. Subsequently, Bob measures the
observable $B_{j}=\sigma_j$ at time $t_{\mathrm{B}}$.

TS can enable Alice to affect on Bob's outcomes. This influence
vanishes when Alice and Bob use uncorrelated bases. In order to
consider only nontrivial TS, we need a channel that provides a
nonunitary evolution of the transmitted qubits. This is because a
pure unitary evolution can be seen as the lack of the evolution of
the transmitted photons in an appropriate reference frame used by
Bob.

The TS inequality of Chen \etal~\cite{Chen14} reads
\begin{equation}\label{eq:ts_ineq}
S_N \equiv \sum_{i=1}^NE\left[ \left\langle
B_{i,t_{\mathrm{B}}}\right\rangle _{A_{i,t_{\mathrm{A}%
}}}^{2}\right] \le 1,
\end{equation}
which is satisfied for all classical states. The TS parameter
$S_N$ that depends on the number $N=2,3$ of unbiased measurements
$B$ performed by Bob. The left-hand-side of the inequality is a
sum over the measurements of the expectation values
$$
E\left[ \left\langle B_{i,t_{\mathrm{B}}}\right\rangle _{A_{i,t_{\mathrm{A}%
}}}^{2}\right] \equiv \sum_{a=\pm 1}P(a|A_{i,t_{\mathrm{A}}})\left\langle B_{i,t_{\mathrm{B}%
}}\right\rangle _{a|A_{i,t_{\mathrm{A}}}}^{2},
$$
where the conditional probability
$$
 P(a|A_{i,t_{\mathrm{A}}})\equiv \sum_{\lambda }q_{\lambda }P_{\lambda
}(a|A_{i,t_{\mathrm{A}}})
$$
depends on a classical variable $\lambda$ that specifies a given
type of channel and $q_{\lambda }$, which specifies the
probability of the qubit being transmitted via the channel
labelled by $\lambda$. The channel here is understood as a single
Kraus operator from the set of the Kraus operators constituting
the map between the state preselected by Alice and that delivered
to Bob. Note that usually Alice and Bob do not know the value of
$\lambda$ (if they did, they could reverse the evolution and
maximize $S_N$) and, thus, this label can be ignored.
Nevertheless, we keep track of this parameter as we change its
value in our experiment. Bob's outcomes are related to the state
projection performed by Alice, as
\begin{equation}\label{eq:B}
\left\langle B_{i,t_{\mathrm{B}}}\right\rangle_{a|A_{i,t_{\mathrm{A}%
}}}\equiv \sum_{b=\pm 1}b\,P(b|A_{i,t_{\mathrm{A}}}=a,B_{i,t_{\mathrm{B}}}%
).
\end{equation}%
The parameter $N$($=2$ or $3$) represents the number of the MUB
used by Bob to analyse the received qubit. The TS
inequality~(\ref{eq:ts_ineq}) follows from the non-commutativity
of two observables that can be measured by Alice and Bob. The
inequality is violated when Alice's choice of observable
influences Bob's result. This could happen only if the channel has
not erased the influence of Alice's choice. It could be said that
temporal steering quantifies the impact of Alice's choice on the
future local reality of Bob. This also means that TS could be used
to witness the causality between time-ordered events. Note that
when the TS parameter reaches its maximal value, it is invariant
with respect to changing the casual relations between Alice and
Bob~\cite{Pawlowski2009,Oreshkov2012,Brukner2014,Procopio2015,Ried2015,Chaves2015}.
In particular, this case could imply the grandfather
paradox:~\cite{Brukner2014} Bob flips his measured state and sends
it backward in time to become the state prepared by Alice.

Breaking the TS inequality corresponds to nonclassical channel
operation causing stronger temporal correlations between Alice and
Bob than the correlations between Alice and the best classical
copy of the transmitted state. The best classical copying strategy
is to measure the state sent by Alice in a random basis and resend
the state further to Bob. The measurement result can be used to
prepare an indefinite number of copies of the state resent to Bob.
In the best case, with the probability $1/N$, the random choice of
basis is compatible with Alice's and Bob's bases and there will be
full correlation between them. However, with the probability
$(N-1)/N$ the choice is not compatible and with probability
$(N-1)/2N$ Alice and Bob receive opposite results. The TS
inequality can be related to the average quantum bit error rate
(QBER) by the following general inequality
\begin{equation}\label{eq:securityUniv}
 \frac{1}{4m_{}}\left(M-\frac{S_N}{N}\right) \geq \mathrm{QBER}_N,
\end{equation}
where $m$ and $M$ are the smallest and largest transmission
fidelity of any state sent by Alice (see Sec.~III), respectively.
For the procedure described above, we substitute $m=1/2$ (which
corresponds to a wrong choice of bases), $M=1$ (for the matched
bases), and $\mathrm{QBER}_N=(N-1)/2N $ to arrive at $S_N\le1$.
This inequality is saturated for the classical copying procedure,
because there are only two possible random values of the fidelity
($1$ and $1/2$). Thus, we derived the TS inequality in a way that
allows to interpret its breaking as a certificate the of lack of
quantum collapse and the occurrence of quantum correlations. This
conclusion holds under the assumption that the sequence of the
measurement bases used by Alice and Bob is secret.

To underline the quantum nature of TS, let us focus on the
security of MUB-based quantum key distribution, which can only be
explained with quantum theory. In this case we can again relate TS
with the average QBER in the raw key obtained by Alice and Bob
after performing the key sifting (basis reconciliation) step in
their MUB protocol (see Sec.~III). This relation reads
$\mathrm{QBER}_N \geq (1-\sqrt{{S_N}/{N}})/2,$ where the
inequality is saturated in the case of isotropic noise. Note that
there is no fundamental reason for considering an anisotropic
noise. In fact, making the noise isotropic is the best strategy
for keeping an eavesdropper undetected. In the MUB protocols,
Alice sends all her basis states to Bob with equal probabilities,
hence there is no preferred direction. For each of the two QKD
protocols there exist a minimal value of $\mathrm{QBER}_{N}=q_N$
for which the respective protocol is no longer secure. When we
consider individual attacks, these values are
$q_2=\tfrac{1}{2}(1-\tfrac{1}{\sqrt{2}})$ for BB84 ($N=2$) and
$q_3=\tfrac{1}{6}$ for B98 ($N=3$). The values of the QBER
correspond to the amount of noise introduced by the respective
optimal isotropic cloning processes, designed to copy the states
sent by Alice. The two cloning regimes are referred to as
phase-covariant and universal cloning for $N=2$ and $N=3$,
respectively. The relation between optimal quantum cloning and the
security of these QKD protocols was studied in various works~(see,
\eg, Refs.~\cite{Gisin02,Soubusta07, Bartuskova07,Lemr12,
Bartkiewicz13prl} and references therein). This connection to
optimal quantum copying is anticipated since the security of the
MUB-based QKD protocols is guaranteed by the impossibility of
ideal copying of an unknown quantum
state~\cite{Wootters82,Dieks82}. {The no-cloning theorem also
explains why it is impossible to send information faster than
light. For any attack, the security condition can then be stated
as
\begin{equation}\label{eq:security_symm}
q_N > \mathrm{QBER}_N\ge
\tfrac{1}{2}\Big(1-\sqrt{\tfrac{S_N}{N}}\Big).
\end{equation}
We can verify this security only if we know a specific value of
QBER${}_N$. The TS parameter $S_N$ is especially useful in the
case of symmetric noise, where $S_N =
4N(\tfrac{1}{2}-\mathrm{QBER}_N)^2$ [or
$\mathrm{QBER_N}=(1-\sqrt{S_N/N})/2$}] and security condition can
be rewritten as
\begin{equation}\label{eq:secI}
S_N>N(1-2q_N)^2\implies\mathrm{QBER}_N<q_N.
\end{equation}
In general, the violation of the security
condition~(\ref{eq:secI}) provides the maximal values of
$S_N=N(1-2q_{N})^2$ for which the relevant QKD protocols are
insecure ($\mathrm{QBER}_N=q_N$). Remarkably, we can show that
Eq.~(\ref{eq:secI}) holds also for asymmetric noise. Note that it
holds $(1-M)\le\mathrm{QBER}_N\le (1-m),$ where $M$ and $m$ are
defined below Eq.~(\ref{eq:securityUniv}). From this it follows
that reaching $\mathrm{QBER}_N\ge q_N,$ which is needed to break
the QKD protocols, requires at least satisfying $m=1-q_N.$ Now,
let us assume that the inequality $S_N>N(1-2q_{N})^2 $ is
satisfied, then $\mathrm{QBER}_N\geq q_N$ or $\mathrm{QBER}_N<q_N$
(which is a trivial case). If  $\mathrm{QBER}_N\geq q_N$, then
from Eq.~(\ref{eq:securityUniv}) for $m=1-q_{N}$ it follows that
$S_N\le N(M-4mq_N)=N(M-2q_N)^2.$ This contradicts the
$S_N>N(1-2q_{N})^2 $ assumption for an arbitrary value of
$M>1-q_N$ (up to $M=1$) and, hence, concludes the proof.

The security condition $S_N>N(1-2q_N)^2$  can be interpreted as a
temporal monogamy relation [corresponding to the violation of the
TS inequality given in Eq.~(\ref{eq:ts_ineq})] for individual
attacks (when $q_N$ is defined by the isotropic optimal cloning)
or collective attacks (when $q_N\propto 0.1$ is defined by an
unconditional security threshold~\cite{secBB84,secB98}),
respectively. The monogamy of temporal correlations is guaranteed
by the secrecy of the sequence of MUBs. If the unconditional
security bounds are reached and even if the environment had access
to all the instances of the experiment simultaneously, then the
environment cannot be better correlated with Alice than Bob
without knowing the sequence of MUBs before Bob's measurements (it
contains his detector but not Alice's setup). This also means that
by breaking the TS inequality [given in Eq.~(\ref{eq:ts_ineq})] by
this proper amount, one certifies the existence of the monogamous
quantum causality, \ie, from the correlations we can infer that
Alice steered Bob's outcomes more than any other party. In this
scenario, the correlation means quantum causation. Moreover, this
causation is monogamous.

The limiting values of the TS parameters are $S_2 = 1$ (at the
TS-inequality threshold) and $S_3 = 4/3>1$ (above the
TS-inequality threshold). Moreover, these values of $S_N$ imply
that the average transfer fidelity is $F_N =
1-\mathrm{QBER}_N>1-q_N$ above the optimal cloning threshold. This
means that the temporal correlations of such strength cannot be
reproduced by probing a single photon sent by Alice in any
physically possible way. From the above analysis it follows that
the violation of the TS inequality is a necessary, but not
sufficient condition for the security of the QKD protocols based
on MUB against individual attacks. The sufficient condition
provided in Eq.~(\ref{eq:secI}) can be interpreted as breaking a
stronger form of the TS inequality, \ie, $S_2\le 1$ and $S_3\le
4/3$ or $S_N\le 2^{N-1}/N$. If this inequality is not broken, then
Alice and Bob should abort their QKD protocol. As found by Chen
\etal~\cite{Chen14}, violating the original $S_2$ inequality
certifies the security of BB84 bounded by the fundamental physical
limitation given by the no-cloning
theorem~\cite{Wootters82,Dieks82}. However, by implementing
coherent attacks (where all the particles sent by Bob are treated
collectively), Eve can induce less noise, which implies a smaller
value of $q_N$ than that for individual attacks (under the
assumption that the sequence of bases used by Alice and Bob is
revealed). We can still use $S_N$ to check the security of the
relevant QKD protocol by applying Eq.~(\ref{eq:secI}). However, we
cannot use it to quantify TS. Analogously to quantifying
nonlocality~\cite{Horst13,Bartkiewicz13PRA}, defining a good
measure of steering is not a simple task. EPR steering has been
studied with specially-designed inequalities leading to
all-versus-nothing measures~\cite{Chen13}. Recently, Skrzypczyk
\etal~\cite{Skrzypczyk14} proposed to quantify EPR steering with a
steerable weight. This measure is described by a semidefinite
program that can be efficiently implemented and provides an
interesting tool for further study of steering. The approach
described by the authors of Ref.~\cite{Skrzypczyk14} is ``allowing
one to explore a wide variety of quantum states and measurement
scenarios.'' However, to our knowledge, a closed formula for the
steerable weight has not been found yet, even for the simplest
scenario. As we show in this paper, this measure can be applied to
experimental data to detect the existence of the TS.

%------------------------------------------------------------------
\section{Temporal steering inequalities and QBER}

Here, we present our theoretical findings relating the temporal
steering inequalities and the security of QKD protocols against
individual attacks for the important case of isotropic noise. Let
us rewrite the steering inequality for the specific case of QKD
protocols, where we assume that $B_i=A_i$. This is granted by the
construction of the QKD protocols, where any other choice of the
observable $B_i$ is not allowed (and it is rejected). The value of
$P(b|A_{i,t_{\mathrm{A}}}=a,B_{i,t_{\mathrm{B}}})$ corresponds to
the conditional probability of Bob measuring the value $b$ under
the condition that his measurement basis is $B_i$ and Alice sent
an eigenstate observable $A_i$ of value $a$. This probability can
be now interpreted as the fidelity $F_i(a,b)$ between the state
measured by Alice and the state delivered to Bob if $b=a$.
Alternatively, the probability equals $1-F_i(a,-b)$ for $b=-a$.
Therefore, Eq.~(\ref{eq:B}) can be rewritten as
$$
\left\langle B_{i,t_{\mathrm{B}}}\right\rangle_{a|A_{i,t_{\mathrm{A}%
}}}\equiv 2F_i(a,1) - 1 = 1-2F_i(a,-1).
$$
Thus,
$$
\left\langle
B_{i,t_{\mathrm{B}}}\right\rangle_{a|A_{i,t_{\mathrm{A}}}}^2 =
(2F_{i,a} - 1)^2,
$$
where $F_{i,a}=F_i(a,b=a)$ is the transmission fidelity of the
eigenstate of $A_i$ associated with the eigenvalue $a$. Now, it
follows from $P(a,b|A_{i,t_{\mathrm{A}}},B_{i,t_{\mathrm{B}}}) =
\frac{1}{2}$, fixed by the construction of the protocols, that the
steering inequality can be rewritten as
$$
S_N \equiv \frac{1}{2}\sum_{i=1}^N \sum_{a=\pm 1} (2F_{i,a} - 1)^2
\leq 1.
$$
It is now apparent that the quantity $S_N$ can be interpreted as
$N$ times the arithmetic mean of $(2F_{i,a} - 1)^2$. This,
quantity can be easily related to the average fidelity using the
Cauchy-Schwartz inequality that implies
$$
\sqrt{N S_N} \geq\sum_{i=1}^N \sum_{a=\pm 1} (F_{i,a} -
\tfrac{1}{2})=2N(F_N-\tfrac{1}{2}),
$$
where $F_N$ is the mean transmission fidelity. In the MUB-based
protocols, the QBER is directly related to the mean fidelity, \ie,
$\mathrm{QBER}_N=1-F_N$. This finally leads to the following
inequality
$$
\mathrm{QBER}_N \geq
\tfrac{1}{2}\left(1-\sqrt{\tfrac{S_N}{N}}\right),
$$
which is saturated if $F_{i,a}=F_N\geq\tfrac{1}{2}$. Using the
expansion of $S_N$ in terms of $F_{i,a}$, and the definition of
$\mathrm{QBER}_N $, we can write
$$
\langle F^2 \rangle = \tfrac{1}{2N}\sum_{i=1}^{N}\sum_{a=\pm1}F_{i,a}^2 = 1-\mathrm{QBER}_N + \frac{S_N-N}{4N}.
$$
Let $\langle (F - F_N)^2 \rangle = \sigma^2$ be the variance of
$F$, where $F_N$ is its mean value. Then, we can express the
variance as
\begin{equation}\label{eq:var}
\sigma^2 = \mathrm{QBER}_N(1-\mathrm{QBER}_N) + \frac{S_N-N}{4N}.
\end{equation}
There exists a strong inequality limiting the values of $\sigma$
from above, \ie, the Barnett-Dragomir~\cite{Barnett99} (or
Bhatia-Davis~\cite{Bhatia00}) inequality for the variable $F$
(which takes, with the same probability, one of the values of
$F_{i,a}$ for $i=1,\,2,\,3$ and $a=\pm1$) that reads
\begin{equation}\label{eq:BDineq}
\sigma^2 \leq (M-F_N)(F_N-m),
\end{equation}
where $M$ is the largest and $m$ the smallest allowed value of
$F_{i,a}$. By substituting $\sigma^2$ in Eq.~(\ref{eq:BDineq})
with the expression given in Eq.~(\ref{eq:var}) we obtain
$$
\mathrm{QBER}_N \leq \tfrac{1}{4m}\left(M-\tfrac{S_N}{N}\right).
$$
This is an upper bound on $\mathrm{QBER}_N$, which is saturated if
$F_N=M$ or $F_N=m$. It is physically justified to use $M=1$ (no
noise) and $m=\tfrac{1}{2}$ (only noise) or $m=\tfrac{3}{4}$ (the
same amounts of signal and noise). A fundamental limit on
transferring information with the particular basis is for
$m=\tfrac{1}{2}$. If it is reached by one of the states, the
number of usable MUB is reduced by one. However, for practical
purposes, the case $m=\tfrac{3}{4}$ is more interesting as it is
related to the quantum limit on fidelity of optimal cloning. This
means that this value is at the quantum threshold of the TS
inequality. The value of $m=\tfrac{3}{4}$ indicates that we do not
use the QKD protocol if any of its basis states is transmitted
with the probability of being randomly flipped to the orthogonal
state larger than $\tfrac{1}{2}$. This refers to the case where a
state is replaced, with probability $\tfrac12$, by a completely
mixed state.
\\

%------------------------------------------------------------------
\section{Temporal steerable weight}

To quantify temporal steering we introduce a direct counterpart of
the EPR steering weight defined by Skrzypczyk
\etal~\cite{Skrzypczyk14} The set of Alice's observables and her
outcomes is given in the form of an assemblage
$\{\rho_{a|A_i}(t_A)\}_{a,i}$. The assemblage encodes the
conditional probability of Alice obtaining the result $a$ when
measuring the observable $A_i$, \ie,
$P(a|A_i)=\Tr[\rho_{a|A_i}(t_A)]$ and the states that are sent to
Bob $\hat{\rho}_{a|A_i}(t_A)= \rho_{a|A_{i,t_{\mathrm{A}}}}(t_A) /
P(a|A_{i,t_{\mathrm{A}}})$. The states received by Bob at time
$t_B$ are altered by a non-unitary channel. Thus, Bob performs his
measurements on the assemblage
$\{\rho_{a|A_i}(t_B)\}_{a,i}\equiv\{\rho_{a|A_i}\}_{a,i}$. A valid
assemblage must satisfy the following consistency relation
\begin{equation}\label{eq:consistency}
\begin{aligned}
    \Tr \sum_{a=\pm1} \rho_{a|A_i} &= 1 \quad\quad \forall i=1,2,3.
\end{aligned}
\end{equation}
The above relations ensure that Bob receives a valid quantum
state.

The unsteerable assemblages, as defined in
Ref.~\cite{Skrzypczyk14}, can be created independently of Alice's
choice of observable (\ie, without entangling Alice's measurement
outcome with the state received by Bob), and can be written in the
following form
\begin{equation} \label{eq:unsteerable}
\begin{aligned}
&\rho_{a|A_i} = \sum_\gamma D_\gamma(a|A_i)\rho_\gamma &\forall a,i, \\
\text{such~that}& \quad\quad \Tr \sum_\gamma \rho_\gamma = 1,
\quad\quad \rho_\gamma \geq 0 &\forall \gamma,
\end{aligned}
\end{equation}
where $\gamma$ is a (classical) random variable held by Alice,
$\rho_\gamma$ are the states received by Bob, and
$D_\gamma(a|A_i)$ are Alice's deterministic functions that map
Alice's variable $\gamma$ to a specific pair of the observable
$A_i$ and outcome $a$. Here we consider only the cases for $N=2,3$
and we list our choices of $D_\gamma(a|A_i)$ in
Tables~\ref{tab:DN2} and \ref{tab:DN3}. Assuming that $N=3$, we
can use Tab.~\ref{tab:DN3} to find that, \eg, $\rho_{+1|A_1} =
\sum_{n=1}^4\rho_n$ or $\rho_{-1|A_1} = \sum_{n=5}^8\rho_n$, etc.
The above-described model of unsteerable assemblage is also known
as the \textit{local hidden state} (LHS) model. The assemblage
that can be described by this model is independent of Alice's
choice of her observable $A_i$, \ie, it is given by
Eq.~\eqref{eq:unsteerable}. {For other (\textit{steerable})
assemblages there is no explanation for how the different
conditional states Bob received could have been prepared by Alice
without her performing the measurements of $A_i$ or sending the
eigenstates of $A_i$. This is why temporal steering is a necessary
condition for implementing the QKD protocols using MUB.

Note that in order to calculate the TS weight $w_{t,2}$ for BB84
(or $w_{t,3}$ for B98), similarly as in the estimation of the QBER
in BB84, Alice has to disclose which bases and Bob has to disclose
his measurement results to Alice. To perform these calculations,
she has to define the assemblage that needs to satisfy the
consistency relation~(\ref{eq:consistency}). The valid assemblage
(just before Bob's measurements) is given by
$\rho_{a|A_i}=\tfrac{1}{2}\hat{\rho}_{a|A_i}(t_B)$. We can rewrite
this assemblage as
\begin{equation*}
\rho_{a|A_i}=\sum_{b,b'=\pm 1} \bra{b,B_i}
\rho_{a|A_i}\ket{b',B_i}\ket{b,B_i}\bra{b',B_i},
\end{equation*}
or, in the special case of $A_i = B_i$
\begin{eqnarray*}\nonumber
\rho_{a|A_i}&=& \sum_{a',a''=\pm 1} \bra{a',A_i} \rho_{a|A_i} \ket{a''|A_i}\ket{a',A_i}\bra{a'',A_i}\\
&=&\frac{1}{2} \sum_{a'=\pm 1} F_{i}(a,a')\oprod{a',
A_i}+\rho_{a'\neq a''}.
\end{eqnarray*}
% This assemblage was used to evaluate the TS weight shown in Fig.~\ref{fig:resultsB}.
The matrix $$\rho_{a'\neq a}=\sum_{a'\neq a''} \bra{a',A_i}
\rho_{a|A_i} \ket{a''|A_i}\ket{a''|A_i}\ket{a',A_i}\bra{a'',A_i} $$
contains correlations that are neglected when calculating the
$S_N$ parameter and it makes the TS weight invariant under any
unitary evolution of the assemblage equivalent to a rotation of
Bob's measurement bases. To obtain this assemblage experimentally,
Bob performs quantum state tomography of the received qubit.}

\begin{table}
\caption{\label{tab:DN2} {Values of deterministic functions
$D_\gamma(a|A)$ and the number of unbiased observables for BB84
($N=2$).} It is shown for each function $D_\gamma(a|A)$ how the
variable $\gamma$ is mapped to a specific set of observables $A$
 and their eigenvalues $a$. }
\begin{ruledtabular}
\begin{tabular}{ccccc}
${a|A}$ & $D_1$ & $D_2$ & $D_3$ & $D_4$ \\
\hline
${-1|A_1}$ & $0$ & $0$ & $1$ & $1$ \\
${+1|A_1}$ & $1$ & $1$ & $0$ & $0$ \\
${-1|A_2}$ & $1$ & $0$ & $1$ & $0$ \\
${+1|A_2}$ & $0$ & $1$ & $0$ & $1$
\end{tabular}
\end{ruledtabular}
\end{table}

\begin{table}
\caption{\label{tab:DN3} {Same as in Table~\ref{tab:DN2}, but
for B98 ($N=3$).}}
\begin{ruledtabular}
\begin{tabular}{ccccccccc}
${a|A}$ & $D_1$ & $D_2$ & $D_3$ & $D_4$ & $D_5$ & $D_6$ & $D_7$ & $D_8$\\
\hline
${-1|A_1}$ & $0$ & $0$ & $0$ & $0$ & $1$ & $1$ & $1$ & $1$\\
${+1|A_1}$ & $1$ & $1$ & $1$ & $1$ & $0$ & $0$ & $0$ & $0$\\
${-1|A_2}$ & $0$ & $0$ & $1$ & $1$ & $0$ & $0$ & $1$ & $1$\\
${+1|A_2}$ & $1$ & $1$ & $0$ & $0$ & $1$ & $1$ & $0$ & $0$\\
${-1|A_3}$ & $0$ & $1$ & $0$ & $1$ & $0$ & $1$ & $0$ & $1$\\
${+1|A_3}$ & $1$ & $0$ & $1$ & $0$ & $1$ & $0$ & $1$ & $0$
\end{tabular}

\end{ruledtabular}
\end{table}

The TS weight $w_t$ is defined as the minimal amount of strictly
steerable resources that is needed to express an arbitrary
assemblage in the following way
\begin{equation}\label{eq:sw_def}
    \rho_{a|A_i} = w_t\rho_{a|A_i}^\S + (1-w_t)\rho_{a|A_i}^\US\quad\quad \forall a,i,
\end{equation}
where $\rho_{a|A_i}^\S$ is a genuine steerable assemblage and
$\rho_{a|A_i}^\US$ is unsteerable [as defined by
Eq.~\eqref{eq:unsteerable}]. The minimum value of $0\leq w_t \leq
1$ for which Eq.~(\ref{eq:sw_def}) is satisfied is the TS weight.
As shown in Ref.~\cite{Skrzypczyk14} the value of $w_t$ is
computed as the solution to the following semidefinite program
(SDP)
\begin{equation*}
\begin{aligned}
\text{max}& \quad \Tr \sum_\gamma \rho_\gamma \\
\text{such~that}& \quad \rho_{a|A_i} - \sum_\gamma D_\gamma(a|A_i)\rho_\gamma \geq 0 \quad &\forall a,i, \\
    & \quad \rho_\gamma \geq 0 \quad &\forall \gamma.
\end{aligned}
\end{equation*}
This SDP can be solved efficiently for small matrices, which is
the case in our experiment. For this purpose we used two SDP
packages which provide consistent results~\cite{SDP1,SDP2,SDP3}.
By analogy with the TS inequality, we might expect that for a
given number of MUB $N$, there exists a value of the TS weight
above which the relevant MUB protocol is secure against individual
attacks~\cite{Chen14,Bartkiewicz13prl}. However, finding the
limiting value of the TS weight for $N=3$ is a difficult task
because of the lack of a closed formula for the TS weight and the
lack of an apparent direct relation between the TS weight and the
TS inequality. Nevertheless, for $N=2$ and uniform noise in the
observables, we can show that the limiting value of the TS weight
is $0$. This is because the \textit{temporal steerability} of an
assemblage can be demonstrated by the violation of the TS
inequality~(\ref{eq:ts_ineq}) or, equivalently, by reaching $w_t >
0$. Thus, for $N=2$ the violation of the TS inequality or reaching
$w_t > 0$ is a necessary and sufficient security condition for the
relevant QKD protocol against individual
attacks.\\

%------------------------------------------------------------------
\section{Discussion and conclusions}

We analysed temporal steering, which is a time-like analog of EPR
steering. The concept of TS is a useful idea that can be applied
to the analysis of QKD protocols. As we showed in this paper, the
TS can be easily observed experimentally, but its relation to
MUB-based QKD protocols is more complex than originally
suspected~\cite{Chen14}. The inequality in
Eq.~(\ref{eq:security_symm}) provides a lower bound on the QBER
related to the TS parameter $S_N$ for cryptographic systems with
arbitrary (isotropic or anisotropic) noise. In the special case of
isotropic noise (which is the case for BB84), we found that the TS
parameter $S_N$ is a simple function of the average transmission
fidelity (or, equivalently, the QBER) and the number of MUB used
in the protocol. We also found an upper bound on the QBER in terms
of $S_N$ given by Eq.~(\ref{eq:securityUniv}). This relation has
also other physical implications beyond testing the security of
the QKD protocols for $m=1/2$. The value of $F_N=(N+1)/(2N)$
corresponds to the classical limit of the fidelity of optimally
copying the evolving state, \ie, splitting the original system
into two equivalently steerable subsystems using such devices that
process only classical information. Thus, if we set
$q_N=1-F_N=(N-1)/(2N)$ in Eq.~(\ref{eq:securityUniv}) and allow an
arbitrary amount of noise ($m=1/2$), then we derive the TS
threshold at $S_N>1$. The TS is a quantum phenomenon, if the
sequence of basis is secret.

There exists a deep relation between the impossibility of
performing perfect quantum cloning and the impossibility of
sending information faster than light. This implies that there are
values of $S_N$ above which it is impossible for Bob to obtain the
outcomes reproducing the correlations before the photon is
physically sent by Alice (reaching his setup) or it is
successfully teleported (so the original one is destroyed). These
values correspond to the quantum-classical cloning threshold,
which implies $S_N\le1$. Reaching $S_N$ above these limiting
values implies that Bob has no access to his future results before
Alice's photon has been successfully delivered. This also means
that Bob witnesses quantum TS because his results can be
counterfeited using only quantum phenomena. Finally, an individual
photon cannot be counterfeited by any quantum process if $S_N>
2^{N-1}/N$. In this case, Bob witnesses the monogamy quantum TS.

There is also another fundamental implication of
Eqs.~(\ref{eq:security_symm}) and (\ref{eq:securityUniv}) as they
explain when we deal with this monogamous quantum TS. If the
security conditions are violated, then temporal correlations can
be induced not only by the particle sent by Bob, but also by, \eg,
its clone. Therefore, this basic requirement on the original
temporal correlations is no longer satisfied as other resources
may lead to the same effect. If, on the other hand, there is no
physically possible way to witness TS without delivering the
original particle, we can be sure that the assumptions used in the
definition of TS are valid. This situation coincides with the
unconditional security threshold~\cite{secBB84, secB98, GisinRMP}
(there exist values of $q_N>0$ for which the protocols are secure)
for MUB-based protocols. This corresponds to $q_N\approx 0.1$ for
$N=2,3$. Note that if the protocols would not be unconditionally
secure ($q_N=0$), then this implies the nonexistence of the
monogamous quantum TS. Thus, we could not distinguish two cases:
whether the same original particle or two different particles
(\eg, the original and its copy) are observed at two different
moments in time. This would mean that the famous phrase of
Heraclitus ``No man ever steps in the same river twice'', taken
literally, could be fundamentally true as there would be now a way
of checking if there exists a single reality of particles evolving
in time. However, the unconditional security of MUB-protocols
shows that  the converse is true because the photons probed in our
experiment display genuine TS. Therefore, the MUB-protocols are
unconditionally secure (against individual attacks) because it is
possible to test whether a particle is the same or not in
time-delayed observations via genuine TS, which reveals stronger
temporal autocorrelations than temporal correlations between
itself and any other particle. This can be referred to as the
monogamy of temporal correlations. Note that all the above
conclusions are valid assuming the sequence of MUBs shared by
Alice and Bob is secret. Our paper introduces the concept of the
monogamy of quantum causality and relates it to TS. Further study
of this concept in the context of casual structures, casual
inference, and the casuality of quantum information could lead to
new fundamental
discoveries~\cite{Pawlowski2009,Oreshkov2012,Brukner2014,Procopio2015,
Ried2015,Chaves2015}. Finally we note that the temporal steerable
weight has recently been described in Ref.~\cite{Chen16}, in
parallel, to this article~\cite{Karol15steering}.

%------------------------------------------------------------------
\section*{Acknowledgments}

The authors thank Shin-Liang Chen, Yueh-Nan Chen, and Neill
Lambert for stimulating discussions. The authors acknowledge
useful discussions with Yueh-Nan Chen, Neill Lambert, and
Shin-Liang Chen on theoretical aspects of temporal steering. K.L.
and K.B. acknowledge the financial support by the Czech Science
Foundation under the project No. 16-10042Y and the financial
support of the Polish National Science Centre under grant
DEC-2013/11/D/ST2/02638. A.\v{C}. acknowledges financial support
by the Czech Science Foundation under the project No.
P205/12/0382. The authors  also acknowledge the project No. LO1305
of the Ministry of Education, Youth and Sports of the Czech
Republic financing the infrastructure of their workplace. A.M. was
supported by the Polish National Science Centre under grants
DEC-2011/03/B/ST2/01903 and DEC-2011/02/A/ST2/00305. F.N. is
partially supported by the RIKEN iTHES Project, MURI Center for
Dynamic Magneto-Optics, JSPS-RFBR contract no. 12-02-92100,
JST-IMPACT, and a Grant-in-Aid for Scientific Research (A).

%------------------------------------------------------------------
%\bibliography{bib_TS}

\begin{thebibliography}{53}%
\makeatletter
\providecommand \@ifxundefined [1]{%
 \@ifx{#1\undefined}
}%
\providecommand \@ifnum [1]{%
 \ifnum #1\expandafter \@firstoftwo
 \else \expandafter \@secondoftwo
 \fi
}%
\providecommand \@ifx [1]{%
 \ifx #1\expandafter \@firstoftwo
 \else \expandafter \@secondoftwo
 \fi
}%
\providecommand \natexlab [1]{#1}%
\providecommand \enquote  [1]{``#1''}%
\providecommand \bibnamefont  [1]{#1}%
\providecommand \bibfnamefont [1]{#1}%
\providecommand \citenamefont [1]{#1}%
\providecommand \href@noop [0]{\@secondoftwo}%
\providecommand \href [0]{\begingroup \@sanitize@url \@href}%
\providecommand \@href[1]{\@@startlink{#1}\@@href}%
\providecommand \@@href[1]{\endgroup#1\@@endlink}%
\providecommand \@sanitize@url [0]{\catcode `\\12\catcode `\$12\catcode
  `\&12\catcode `\#12\catcode `\^12\catcode `\_12\catcode `\%12\relax}%
\providecommand \@@startlink[1]{}%
\providecommand \@@endlink[0]{}%
\providecommand \url  [0]{\begingroup\@sanitize@url \@url }%
\providecommand \@url [1]{\endgroup\@href {#1}{\urlprefix }}%
\providecommand \urlprefix  [0]{URL }%
\providecommand \Eprint [0]{\href }%
\providecommand \doibase [0]{http://dx.doi.org/}%
\providecommand \selectlanguage [0]{\@gobble}%
\providecommand \bibinfo  [0]{\@secondoftwo}%
\providecommand \bibfield  [0]{\@secondoftwo}%
\providecommand \translation [1]{[#1]}%
\providecommand \BibitemOpen [0]{}%
\providecommand \bibitemStop [0]{}%
\providecommand \bibitemNoStop [0]{.\EOS\space}%
\providecommand \EOS [0]{\spacefactor3000\relax}%
\providecommand \BibitemShut  [1]{\csname bibitem#1\endcsname}%
\let\auto@bib@innerbib\@empty
%</preamble>
\bibitem [{\citenamefont {Schr\"{o}dinger}(1935)}]{Schrodinger35}%
  \BibitemOpen
  \bibfield  {author} {\bibinfo {author} {\bibfnamefont {E.}~\bibnamefont
  {Schr\"{o}dinger}},\ }\bibfield  {title} {\enquote {\bibinfo {title}
  {Discussion of probability relations between separated systems},}\ }\href
  {\doibase 10.1017/S0305004100013554} {\bibfield  {journal} {\bibinfo
  {journal} {Math. Proc. Camb. Phil. Soc.}\ }\textbf {\bibinfo {volume} {31}},\
  \bibinfo {pages} {555--563} (\bibinfo {year} {1935})}\BibitemShut {NoStop}%
\bibitem [{\citenamefont {Einstein}\ \emph {et~al.}(1935)\citenamefont
  {Einstein}, \citenamefont {Podolsky},\ and\ \citenamefont {Rosen}}]{EPR35}%
  \BibitemOpen
  \bibfield  {author} {\bibinfo {author} {\bibfnamefont {A.}~\bibnamefont
  {Einstein}}, \bibinfo {author} {\bibfnamefont {B.}~\bibnamefont {Podolsky}},
  \ and\ \bibinfo {author} {\bibfnamefont {N.}~\bibnamefont {Rosen}},\
  }\bibfield  {title} {\enquote {\bibinfo {title} {Can quantum-mechanical
  description of physical reality be considered complete?}}\ }\href {\doibase
  10.1103/PhysRev.47.777} {\bibfield  {journal} {\bibinfo  {journal} {Phys.
  Rev.}\ }\textbf {\bibinfo {volume} {47}},\ \bibinfo {pages} {777--780}
  (\bibinfo {year} {1935})}\BibitemShut {NoStop}%
\bibitem [{EPR(2015)}]{EPR80}%
  \BibitemOpen
  \bibfield  {title} {\enquote {\bibinfo {title} {The special issue of {J. Opt.
  Soc. B }on ``80 years of steering and the {E}instein--{P}odolsky--{R}osen
  paradox''},}\ }\href {\doibase 10.1364/JOSAB.32.00EPR1} {\bibfield  {journal}
  {\bibinfo  {journal} {J. Opt. Soc. B}\ }\textbf {\bibinfo {volume} {32}},\
  \bibinfo {pages} {A1--A91} (\bibinfo {year} {2015})}\BibitemShut {NoStop}%
\bibitem [{\citenamefont {Saunders}\ \emph {et~al.}(2010)\citenamefont
  {Saunders}, \citenamefont {Jones}, \citenamefont {Wiseman},\ and\
  \citenamefont {Pryde}}]{Saunders10}%
  \BibitemOpen
  \bibfield  {author} {\bibinfo {author} {\bibfnamefont {D.~J.}\ \bibnamefont
  {Saunders}}, \bibinfo {author} {\bibfnamefont {S.~J.}\ \bibnamefont {Jones}},
  \bibinfo {author} {\bibfnamefont {H.~M.}\ \bibnamefont {Wiseman}}, \ and\
  \bibinfo {author} {\bibfnamefont {G.~J.}\ \bibnamefont {Pryde}},\ }\bibfield
  {title} {\enquote {\bibinfo {title} {Experimental {EPR}-steering using
  {B}ell-local states},}\ }\href@noop {} {\bibfield  {journal} {\bibinfo
  {journal} {Nat. Phys.}\ }\textbf {\bibinfo {volume} {6}},\ \bibinfo {pages}
  {845--849} (\bibinfo {year} {2010})}\BibitemShut {NoStop}%
\bibitem [{\citenamefont {Walborn}\ \emph {et~al.}(2011)\citenamefont
  {Walborn}, \citenamefont {Salles}, \citenamefont {Gomes}, \citenamefont
  {Toscano},\ and\ \citenamefont {Souto~Ribeiro}}]{Walborn11}%
  \BibitemOpen
  \bibfield  {author} {\bibinfo {author} {\bibfnamefont {S.~P.}\ \bibnamefont
  {Walborn}}, \bibinfo {author} {\bibfnamefont {A.}~\bibnamefont {Salles}},
  \bibinfo {author} {\bibfnamefont {R.~M.}\ \bibnamefont {Gomes}}, \bibinfo
  {author} {\bibfnamefont {F.}~\bibnamefont {Toscano}}, \ and\ \bibinfo
  {author} {\bibfnamefont {P.~H.}\ \bibnamefont {Souto~Ribeiro}},\ }\bibfield
  {title} {\enquote {\bibinfo {title} {Revealing hidden
  {E}instein-{P}odolsky-{R}osen nonlocality},}\ }\href {\doibase
  10.1103/PhysRevLett.106.130402} {\bibfield  {journal} {\bibinfo  {journal}
  {Phys. Rev. Lett.}\ }\textbf {\bibinfo {volume} {106}},\ \bibinfo {pages}
  {130402} (\bibinfo {year} {2011})}\BibitemShut {NoStop}%
\bibitem [{\citenamefont {Wittmann}\ \emph {et~al.}(2012)\citenamefont
  {Wittmann}, \citenamefont {Ramelow}, \citenamefont {Steinlechner},
  \citenamefont {Langford}, \citenamefont {Brunner}, \citenamefont {Wiseman},
  \citenamefont {Ursin},\ and\ \citenamefont {Zeilinger}}]{Wittmann12}%
  \BibitemOpen
  \bibfield  {author} {\bibinfo {author} {\bibfnamefont {B.}~\bibnamefont
  {Wittmann}}, \bibinfo {author} {\bibfnamefont {S.}~\bibnamefont {Ramelow}},
  \bibinfo {author} {\bibfnamefont {F.}~\bibnamefont {Steinlechner}}, \bibinfo
  {author} {\bibfnamefont {N.~K.}\ \bibnamefont {Langford}}, \bibinfo {author}
  {\bibfnamefont {N.}~\bibnamefont {Brunner}}, \bibinfo {author} {\bibfnamefont
  {H.~M.}\ \bibnamefont {Wiseman}}, \bibinfo {author} {\bibfnamefont
  {R.}~\bibnamefont {Ursin}}, \ and\ \bibinfo {author} {\bibfnamefont
  {A.}~\bibnamefont {Zeilinger}},\ }\bibfield  {title} {\enquote {\bibinfo
  {title} {Loophole-free {E}instein-{P}odolsky-{R}osen experiment via quantum
  steering},}\ }\href@noop {} {\bibfield  {journal} {\bibinfo  {journal} {New
  J. Phys.}\ }\textbf {\bibinfo {volume} {14}},\ \bibinfo {pages} {053030}
  (\bibinfo {year} {2012})}\BibitemShut {NoStop}%
\bibitem [{\citenamefont {Smith}\ \emph {et~al.}(2012)\citenamefont {Smith},
  \citenamefont {Gillett}, \citenamefont {de~Almeida}, \citenamefont
  {Branciard}, \citenamefont {Fedrizzi}, \citenamefont {Weinhold},
  \citenamefont {Lita}, \citenamefont {Calkins}, \citenamefont {Gerrits},
  \citenamefont {Wiseman}, \citenamefont {Nam},\ and\ \citenamefont
  {White}}]{Smith12}%
  \BibitemOpen
  \bibfield  {author} {\bibinfo {author} {\bibfnamefont {D.~H.}\ \bibnamefont
  {Smith}}, \bibinfo {author} {\bibfnamefont {G.}~\bibnamefont {Gillett}},
  \bibinfo {author} {\bibfnamefont {M.~P.}\ \bibnamefont {de~Almeida}},
  \bibinfo {author} {\bibfnamefont {C.}~\bibnamefont {Branciard}}, \bibinfo
  {author} {\bibfnamefont {A.}~\bibnamefont {Fedrizzi}}, \bibinfo {author}
  {\bibfnamefont {T.~J.}\ \bibnamefont {Weinhold}}, \bibinfo {author}
  {\bibfnamefont {A.}~\bibnamefont {Lita}}, \bibinfo {author} {\bibfnamefont
  {B.}~\bibnamefont {Calkins}}, \bibinfo {author} {\bibfnamefont
  {T.}~\bibnamefont {Gerrits}}, \bibinfo {author} {\bibfnamefont {H.~M.}\
  \bibnamefont {Wiseman}}, \bibinfo {author} {\bibfnamefont {S.~W.}\
  \bibnamefont {Nam}}, \ and\ \bibinfo {author} {\bibfnamefont {A.~G.}\
  \bibnamefont {White}},\ }\bibfield  {title} {\enquote {\bibinfo {title}
  {Conclusive quantum steering with superconducting transition-edge sensors},}\
  }\href {\doibase http://dx.doi.org/10.1038/ncomms1628} {\bibfield  {journal}
  {\bibinfo  {journal} {Nat. Commun.}\ }\textbf {\bibinfo {volume} {3}},\
  \bibinfo {pages} {625} (\bibinfo {year} {2012})}\BibitemShut {NoStop}%
\bibitem [{\citenamefont {Bennet}\ \emph {et~al.}(2012)\citenamefont {Bennet},
  \citenamefont {Evans}, \citenamefont {Saunders}, \citenamefont {Branciard},
  \citenamefont {Cavalcanti}, \citenamefont {Wiseman},\ and\ \citenamefont
  {Pryde}}]{Bennet12}%
  \BibitemOpen
  \bibfield  {author} {\bibinfo {author} {\bibfnamefont {A.~J.}\ \bibnamefont
  {Bennet}}, \bibinfo {author} {\bibfnamefont {D.~A.}\ \bibnamefont {Evans}},
  \bibinfo {author} {\bibfnamefont {D.~J.}\ \bibnamefont {Saunders}}, \bibinfo
  {author} {\bibfnamefont {C.}~\bibnamefont {Branciard}}, \bibinfo {author}
  {\bibfnamefont {E.~G.}\ \bibnamefont {Cavalcanti}}, \bibinfo {author}
  {\bibfnamefont {H.~M.}\ \bibnamefont {Wiseman}}, \ and\ \bibinfo {author}
  {\bibfnamefont {G.~J.}\ \bibnamefont {Pryde}},\ }\bibfield  {title} {\enquote
  {\bibinfo {title} {Arbitrarily loss-tolerant {E}instein-{P}odolsky-{R}osen
  steering allowing a demonstration over 1~km of optical fiber with no
  detection loophole},}\ }\href {\doibase 10.1103/PhysRevX.2.031003} {\bibfield
   {journal} {\bibinfo  {journal} {Phys. Rev. X}\ }\textbf {\bibinfo {volume}
  {2}},\ \bibinfo {pages} {031003} (\bibinfo {year} {2012})}\BibitemShut
  {NoStop}%
\bibitem [{\citenamefont {H{\"a}ndchen}\ \emph {et~al.}(2012)\citenamefont
  {H{\"a}ndchen}, \citenamefont {Eberle}, \citenamefont {Steinlechner},
  \citenamefont {Samblowski}, \citenamefont {Franz}, \citenamefont {Werner},\
  and\ \citenamefont {Schnabel}}]{Handchen12}%
  \BibitemOpen
  \bibfield  {author} {\bibinfo {author} {\bibfnamefont {V.}~\bibnamefont
  {H{\"a}ndchen}}, \bibinfo {author} {\bibfnamefont {T.}~\bibnamefont
  {Eberle}}, \bibinfo {author} {\bibfnamefont {S.}~\bibnamefont
  {Steinlechner}}, \bibinfo {author} {\bibfnamefont {A.}~\bibnamefont
  {Samblowski}}, \bibinfo {author} {\bibfnamefont {T.}~\bibnamefont {Franz}},
  \bibinfo {author} {\bibfnamefont {R.~F.}\ \bibnamefont {Werner}}, \ and\
  \bibinfo {author} {\bibfnamefont {R.}~\bibnamefont {Schnabel}},\ }\bibfield
  {title} {\enquote {\bibinfo {title} {Observation of one-way
  {E}instein-{P}odolsky-{R}osen steering},}\ }\href@noop {} {\bibfield
  {journal} {\bibinfo  {journal} {Nat. Photon.}\ }\textbf {\bibinfo {volume}
  {6}},\ \bibinfo {pages} {596--599} (\bibinfo {year} {2012})}\BibitemShut
  {NoStop}%
\bibitem [{\citenamefont {Steinlechner}\ \emph {et~al.}(2013)\citenamefont
  {Steinlechner}, \citenamefont {Bauchrowitz}, \citenamefont {Eberle},\ and\
  \citenamefont {Schnabel}}]{Steinlechner13}%
  \BibitemOpen
  \bibfield  {author} {\bibinfo {author} {\bibfnamefont {S.}~\bibnamefont
  {Steinlechner}}, \bibinfo {author} {\bibfnamefont {J.}\ \bibnamefont
  {Bauchrowitz}}, \bibinfo {author} {\bibfnamefont {T.}~\bibnamefont {Eberle}},
  \ and\ \bibinfo {author} {\bibfnamefont {R.}~\bibnamefont {Schnabel}},\
  }\bibfield  {title} {\enquote {\bibinfo {title} {Strong
  {E}instein-{P}odolsky-{R}osen steering with unconditional entangled
  states},}\ }\href {\doibase 10.1103/PhysRevA.87.022104} {\bibfield  {journal}
  {\bibinfo  {journal} {Phys. Rev. A}\ }\textbf {\bibinfo {volume} {87}},\
  \bibinfo {pages} {022104} (\bibinfo {year} {2013})}\BibitemShut {NoStop}%
\bibitem [{\citenamefont {Su}\ \emph {et~al.}(2013)\citenamefont {Su},
  \citenamefont {Chen}, \citenamefont {Wu}, \citenamefont {Deng},\ and\
  \citenamefont {Oh}}]{Su13}%
  \BibitemOpen
  \bibfield  {author} {\bibinfo {author} {\bibfnamefont {H.~Y.}\ \bibnamefont
  {Su}}, \bibinfo {author} {\bibfnamefont {J.~L.}\ \bibnamefont {Chen}},
  \bibinfo {author} {\bibfnamefont {C.}~\bibnamefont {Wu}}, \bibinfo {author}
  {\bibfnamefont {D.~L.}\ \bibnamefont {Deng}}, \ and\ \bibinfo {author}
  {\bibfnamefont {C.~H.}\ \bibnamefont {Oh}},\ }\bibfield  {title} {\enquote
  {\bibinfo {title} {Detecting {E}instein-{P}odolsky-{R}osen steering for
  continuous variable wavefunctions},}\ }\href {\doibase
  10.1142/S0219749913500196} {\bibfield  {journal} {\bibinfo  {journal} {I. J.
  Quant. Infor.}\ }\textbf {\bibinfo {volume} {11}},\ \bibinfo {pages}
  {1350019} (\bibinfo {year} {2013})}\BibitemShut {NoStop}%
\bibitem [{\citenamefont {Schneeloch}\ \emph {et~al.}(2013)\citenamefont
  {Schneeloch}, \citenamefont {Dixon}, \citenamefont {Howland}, \citenamefont
  {Broadbent},\ and\ \citenamefont {Howell}}]{Schneeloch13}%
  \BibitemOpen
  \bibfield  {author} {\bibinfo {author} {\bibfnamefont {J.}~\bibnamefont
  {Schneeloch}}, \bibinfo {author} {\bibfnamefont {P.~B.}\ \bibnamefont
  {Dixon}}, \bibinfo {author} {\bibfnamefont {G.~A.}\ \bibnamefont {Howland}},
  \bibinfo {author} {\bibfnamefont {C.~J.}\ \bibnamefont {Broadbent}}, \ and\
  \bibinfo {author} {\bibfnamefont {J.~C.}\ \bibnamefont {Howell}},\ }\bibfield
   {title} {\enquote {\bibinfo {title} {Violation of continuous-variable
  {E}instein-{P}odolsky-{R}osen steering with discrete measurements},}\ }\href
  {\doibase 10.1103/PhysRevLett.110.130407} {\bibfield  {journal} {\bibinfo
  {journal} {Phys. Rev. Lett.}\ }\textbf {\bibinfo {volume} {110}},\ \bibinfo
  {pages} {130407} (\bibinfo {year} {2013})}\BibitemShut {NoStop}%
\bibitem [{\citenamefont {He}\ \emph {et~al.}(2015)\citenamefont {He},
  \citenamefont {Rosales-Z\'arate}, \citenamefont {Adesso},\ and\ \citenamefont
  {Reid}}]{He15}%
  \BibitemOpen
  \bibfield  {author} {\bibinfo {author} {\bibfnamefont {Q.}~\bibnamefont
  {He}}, \bibinfo {author} {\bibfnamefont {L.}~\bibnamefont
  {Rosales-Z\'arate}}, \bibinfo {author} {\bibfnamefont {G.}~\bibnamefont
  {Adesso}}, \ and\ \bibinfo {author} {\bibfnamefont {M.~D.}\ \bibnamefont
  {Reid}},\ }\bibfield  {title} {\enquote {\bibinfo {title} {Secure
  {C}ontinuous {V}ariable {T}eleportation and {E}instein-{P}odolsky-{R}osen
  {S}teering},}\ }\href {\doibase 10.1103/PhysRevLett.115.180502} {\bibfield
  {journal} {\bibinfo  {journal} {Phys. Rev. Lett.}\ }\textbf {\bibinfo
  {volume} {115}},\ \bibinfo {pages} {180502} (\bibinfo {year}
  {2015})}\BibitemShut {NoStop}%
\bibitem [{\citenamefont {Pusey}(2015)}]{Pusey15}%
  \BibitemOpen
  \bibfield  {author} {\bibinfo {author} {\bibfnamefont {M.~F.}\ \bibnamefont
  {Pusey}},\ }\bibfield  {title} {\enquote {\bibinfo {title} {Verifying the
  quantumness of a channel with an untrusted device},}\ }\href@noop {}
  {\bibfield  {journal} {\bibinfo  {journal} {J. Opt. Soc. Am. B}\ }\textbf
  {\bibinfo {volume} {32}},\ \bibinfo {pages} {A56} (\bibinfo {year}
  {2015})}\BibitemShut {NoStop}%
\bibitem [{\citenamefont {Kocsis}\ \emph {et~al.}(2015)\citenamefont {Kocsis},
  \citenamefont {Hall}, \citenamefont {Bennet}, \citenamefont {Saunders},\ and\
  \citenamefont {Pryde}}]{Kocsis15}%
  \BibitemOpen
  \bibfield  {author} {\bibinfo {author} {\bibfnamefont {S.}~\bibnamefont
  {Kocsis}}, \bibinfo {author} {\bibfnamefont {M.~J.~W.}\ \bibnamefont {Hall}},
  \bibinfo {author} {\bibfnamefont {A.~J.}\ \bibnamefont {Bennet}}, \bibinfo
  {author} {\bibfnamefont {D.~J.}\ \bibnamefont {Saunders}}, \ and\ \bibinfo
  {author} {\bibfnamefont {G.~J.}\ \bibnamefont {Pryde}},\ }\bibfield  {title}
  {\enquote {\bibinfo {title} {Experimental measurement-device-independent
  verification of quantum steering},}\ }\href@noop {} {\bibfield  {journal}
  {\bibinfo  {journal} {Nat. Commun.}\ }\textbf {\bibinfo {volume} {6}}
  (\bibinfo {year} {2015})}\BibitemShut {NoStop}%
\bibitem [{\citenamefont {Chen}\ \emph {et~al.}(2014)\citenamefont {Chen},
  \citenamefont {Li}, \citenamefont {Lambert}, \citenamefont {Chen},
  \citenamefont {Ota}, \citenamefont {Chen},\ and\ \citenamefont
  {Nori}}]{Chen14}%
  \BibitemOpen
  \bibfield  {author} {\bibinfo {author} {\bibfnamefont {Y.-N.}\ \bibnamefont
  {Chen}}, \bibinfo {author} {\bibfnamefont {C.-M.}\ \bibnamefont {Li}},
  \bibinfo {author} {\bibfnamefont {N.}~\bibnamefont {Lambert}}, \bibinfo
  {author} {\bibfnamefont {S.-L.}\ \bibnamefont {Chen}}, \bibinfo {author}
  {\bibfnamefont {Y.}~\bibnamefont {Ota}}, \bibinfo {author} {\bibfnamefont
  {G.-Y.}\ \bibnamefont {Chen}}, \ and\ \bibinfo {author} {\bibfnamefont
  {F.}~\bibnamefont {Nori}},\ }\bibfield  {title} {\enquote {\bibinfo {title}
  {Temporal steering inequality},}\ }\href {\doibase
  10.1103/PhysRevA.89.032112} {\bibfield  {journal} {\bibinfo  {journal} {Phys.
  Rev. A}\ }\textbf {\bibinfo {volume} {89}},\ \bibinfo {pages} {032112}
  (\bibinfo {year} {2014})}\BibitemShut {NoStop}%
\bibitem [{\citenamefont {Li}\ \emph {et~al.}(2015)\citenamefont {Li},
  \citenamefont {Chen}, \citenamefont {Lambert}, \citenamefont {Chiu},\ and\
  \citenamefont {Nori}}]{Li15}%
  \BibitemOpen
  \bibfield  {author} {\bibinfo {author} {\bibfnamefont {Che-Ming}\
  \bibnamefont {Li}}, \bibinfo {author} {\bibfnamefont {Yueh-Nan}\ \bibnamefont
  {Chen}}, \bibinfo {author} {\bibfnamefont {Neill}\ \bibnamefont {Lambert}},
  \bibinfo {author} {\bibfnamefont {Ching-Yi}\ \bibnamefont {Chiu}}, \ and\
  \bibinfo {author} {\bibfnamefont {Franco}\ \bibnamefont {Nori}},\ }\bibfield
  {title} {\enquote {\bibinfo {title} {Certifying single-system steering for
  quantum-information processing},}\ }\href {\doibase
  10.1103/PhysRevA.92.062310} {\bibfield  {journal} {\bibinfo  {journal} {Phys.
  Rev. A}\ }\textbf {\bibinfo {volume} {92}},\ \bibinfo {pages} {062310}
  (\bibinfo {year} {2015})}\BibitemShut {NoStop}%
\bibitem [{\citenamefont {Karthik}\ \emph {et~al.}(2015)\citenamefont
  {Karthik}, \citenamefont {Tej}, \citenamefont {Devi},\ and\ \citenamefont
  {Rajagopal}}]{Karthik15}%
  \BibitemOpen
  \bibfield  {author} {\bibinfo {author} {\bibfnamefont {H.~S.}\ \bibnamefont
  {Karthik}}, \bibinfo {author} {\bibfnamefont {J.~Prabhu}\ \bibnamefont
  {Tej}}, \bibinfo {author} {\bibfnamefont {A.~R.~Usha}\ \bibnamefont {Devi}},
  \ and\ \bibinfo {author} {\bibfnamefont {A.~K.}\ \bibnamefont {Rajagopal}},\
  }\bibfield  {title} {\enquote {\bibinfo {title} {Joint measurability and
  temporal steering},}\ }\href {\doibase 10.1364/JOSAB.32.000A34} {\bibfield
  {journal} {\bibinfo  {journal} {J. Opt. Soc. Am. B}\ }\textbf {\bibinfo
  {volume} {32}},\ \bibinfo {pages} {A34--A39} (\bibinfo {year}
  {2015})}\BibitemShut {NoStop}%
\bibitem [{\citenamefont {Mal}\ \emph {et~al.}(2015)\citenamefont {Mal},
  \citenamefont {Majumdar},\ and\ \citenamefont {Home}}]{Mal15}%
  \BibitemOpen
  \bibfield  {author} {\bibinfo {author} {\bibfnamefont {S.}~\bibnamefont
  {Mal}}, \bibinfo {author} {\bibfnamefont {A.~S.}\ \bibnamefont {Majumdar}}, \
  and\ \bibinfo {author} {\bibfnamefont {D.}~\bibnamefont {Home}},\ }\href
  {http://arxiv.org/abs/1510.00625} {\enquote {\bibinfo {title} {Hierarchy of
  temporal correlations in quantum mechanics},}\ } (\bibinfo {year} {2015}),\
  \bibinfo {note} {arXiv:1510.00625 [quant-ph]}\BibitemShut {NoStop}%
\bibitem [{\citenamefont {Chen}\ \emph {et~al.}(2015)\citenamefont {Chen},
  \citenamefont {Chao},\ and\ \citenamefont {Chen}}]{Chen15}%
  \BibitemOpen
  \bibfield  {author} {\bibinfo {author} {\bibfnamefont {S.-L.}\ \bibnamefont
  {Chen}}, \bibinfo {author} {\bibfnamefont {C.-S.}\ \bibnamefont {Chao}}, \
  and\ \bibinfo {author} {\bibfnamefont {Y.-N.}\ \bibnamefont {Chen}},\
  }\bibfield  {title} {\enquote {\bibinfo {title} {Detecting the existence of
  an invisibility cloak using temporal steering},}\ }\href@noop {} {\bibfield
  {journal} {\bibinfo  {journal} {Sci. Rep.}\ }\textbf {\bibinfo {volume} {5}}
  (\bibinfo {year} {2015})}\BibitemShut {NoStop}%
\bibitem [{\citenamefont {Chen}\ \emph {et~al.}(2016)\citenamefont {Chen},
  \citenamefont {Lambert}, \citenamefont {Li}, \citenamefont {Miranowicz},
  \citenamefont {Chen},\ and\ \citenamefont {Nori}}]{Chen16}%
  \BibitemOpen
  \bibfield  {author} {\bibinfo {author} {\bibfnamefont {Sh.~L.}\ \bibnamefont
  {Chen}}, \bibinfo {author} {\bibfnamefont {N.}~\bibnamefont {Lambert}},
  \bibinfo {author} {\bibfnamefont {Ch.~M.}\ \bibnamefont {Li}}, \bibinfo
  {author} {\bibfnamefont {A.}~\bibnamefont {Miranowicz}}, \bibinfo {author}
  {\bibfnamefont {Y.~N.}\ \bibnamefont {Chen}}, \ and\ \bibinfo {author}
  {\bibfnamefont {F.}~\bibnamefont {Nori}},\ }\bibfield  {title} {\enquote
  {\bibinfo {title} {Quantifying non-markovianity with temporal steering},}\
  }\href {\doibase 10.1103/PhysRevLett.116.020503} {\bibfield  {journal}
  {\bibinfo  {journal} {Phys. Rev. Lett.}\ }\textbf {\bibinfo {volume} {116}},\
  \bibinfo {pages} {020503} (\bibinfo {year} {2016})}\BibitemShut {NoStop}%
\bibitem [{\citenamefont {Chiu}\ \emph {et~al.}(2016)\citenamefont {Chiu},
  \citenamefont {Lambert}, \citenamefont {Liao}, \citenamefont {Nori},\ and\
  \citenamefont {Li}}]{Chiu16}%
  \BibitemOpen
  \bibfield  {author} {\bibinfo {author} {\bibfnamefont {C.-Y.}\ \bibnamefont
  {Chiu}}, \bibinfo {author} {\bibfnamefont {N.}~\bibnamefont {Lambert}},
  \bibinfo {author} {\bibfnamefont {T.-L.}\ \bibnamefont {Liao}}, \bibinfo
  {author} {\bibfnamefont {F.}~\bibnamefont {Nori}}, \ and\ \bibinfo {author}
  {\bibfnamefont {C.-M.}\ \bibnamefont {Li}},\ }\href
  {http://arxiv.org/abs/1601.04407} {\enquote {\bibinfo {title} {No-cloning of
  quantum steering},}\ } (\bibinfo {year} {2016}),\ \bibinfo {note}
  {arXiv:1601.04407 [quant-ph]}\BibitemShut {NoStop}%
\bibitem [{Pla()}]{Plato}%
  \BibitemOpen
  \href@noop {} {}\bibinfo {note} {This is how Plato puts the Heraclitus
  doctrine. See Plato's {\em Cratylus}, 402a.}\BibitemShut {Stop}%
\bibitem [{\citenamefont {Benett}\ and\ \citenamefont {Brassard}(1984)}]{BB84}%
  \BibitemOpen
  \bibfield  {author} {\bibinfo {author} {\bibfnamefont {C.H.}\ \bibnamefont
  {Benett}}\ and\ \bibinfo {author} {\bibfnamefont {G.}~\bibnamefont
  {Brassard}},\ }\bibfield  {title} {\enquote {\bibinfo {title} {Public key
  distribution and coin tossing},}\ }\href@noop {} {\bibfield  {journal}
  {\bibinfo  {journal} {In Proc. IEEE Int. Conf. on Computers, Systems and
  Signal Processing}\ }\textbf {\bibinfo {volume} {175}},\ \bibinfo {pages} {8}
  (\bibinfo {year} {1984})}\BibitemShut {NoStop}%
\bibitem [{\citenamefont {Bru\ss{}}(1998)}]{Bruss98}%
  \BibitemOpen
  \bibfield  {author} {\bibinfo {author} {\bibfnamefont {D.}~\bibnamefont
  {Bru\ss{}}},\ }\bibfield  {title} {\enquote {\bibinfo {title} {Optimal
  eavesdropping in quantum cryptography with six states},}\ }\href {\doibase
  10.1103/PhysRevLett.81.3018} {\bibfield  {journal} {\bibinfo  {journal}
  {Phys. Rev. Lett.}\ }\textbf {\bibinfo {volume} {81}},\ \bibinfo {pages}
  {3018--3021} (\bibinfo {year} {1998})}\BibitemShut {NoStop}%
\bibitem [{\citenamefont {Shor}\ and\ \citenamefont
  {Preskill}(2000)}]{secBB84}%
  \BibitemOpen
  \bibfield  {author} {\bibinfo {author} {\bibfnamefont {P.~W.}\ \bibnamefont
  {Shor}}\ and\ \bibinfo {author} {\bibfnamefont {J.}~\bibnamefont
  {Preskill}},\ }\bibfield  {title} {\enquote {\bibinfo {title} {Simple proof
  of security of the {BB84} quantum key distribution protocol},}\ }\href
  {\doibase 10.1103/PhysRevLett.85.441} {\bibfield  {journal} {\bibinfo
  {journal} {Phys. Rev. Lett.}\ }\textbf {\bibinfo {volume} {85}},\ \bibinfo
  {pages} {441--444} (\bibinfo {year} {2000})}\BibitemShut {NoStop}%
\bibitem [{\citenamefont {Lo}(2001)}]{secB98}%
  \BibitemOpen
  \bibfield  {author} {\bibinfo {author} {\bibfnamefont {H.-K.}\ \bibnamefont
  {Lo}},\ }\bibfield  {title} {\enquote {\bibinfo {title} {Proof of
  unconditional security of six-state quatum key distribution scheme},}\
  }\href@noop {} {\bibfield  {journal} {\bibinfo  {journal} {Quantum
  Information and Computation}\ }\textbf {\bibinfo {volume} {1}},\ \bibinfo
  {pages} {81--94} (\bibinfo {year} {2001})}\BibitemShut {NoStop}%
\bibitem [{\citenamefont {Gisin}\ \emph
  {et~al.}(2002{\natexlab{a}})\citenamefont {Gisin}, \citenamefont {Ribordy},
  \citenamefont {Tittel},\ and\ \citenamefont {Zbinden}}]{GisinRMP}%
  \BibitemOpen
  \bibfield  {author} {\bibinfo {author} {\bibfnamefont {N.}~\bibnamefont
  {Gisin}}, \bibinfo {author} {\bibfnamefont {G.}~\bibnamefont {Ribordy}},
  \bibinfo {author} {\bibfnamefont {W.}~\bibnamefont {Tittel}}, \ and\ \bibinfo
  {author} {\bibfnamefont {H.}~\bibnamefont {Zbinden}},\ }\bibfield  {title}
  {\enquote {\bibinfo {title} {Quantum cryptography},}\ }\href {\doibase
  10.1103/RevModPhys.74.145} {\bibfield  {journal} {\bibinfo  {journal} {Rev.
  Mod. Phys.}\ }\textbf {\bibinfo {volume} {74}},\ \bibinfo {pages} {145--195}
  (\bibinfo {year} {2002}{\natexlab{a}})}\BibitemShut {NoStop}%
\bibitem [{\citenamefont {Skrzypczyk}\ \emph {et~al.}(2014)\citenamefont
  {Skrzypczyk}, \citenamefont {Navascu\'es},\ and\ \citenamefont
  {Cavalcanti}}]{Skrzypczyk14}%
  \BibitemOpen
  \bibfield  {author} {\bibinfo {author} {\bibfnamefont {P.}~\bibnamefont
  {Skrzypczyk}}, \bibinfo {author} {\bibfnamefont {M.}~\bibnamefont
  {Navascu\'es}}, \ and\ \bibinfo {author} {\bibfnamefont {D.}~\bibnamefont
  {Cavalcanti}},\ }\bibfield  {title} {\enquote {\bibinfo {title} {Quantifying
  {E}instein-{P}odolsky-{R}osen steering},}\ }\href {\doibase
  10.1103/PhysRevLett.112.180404} {\bibfield  {journal} {\bibinfo  {journal}
  {Phys. Rev. Lett.}\ }\textbf {\bibinfo {volume} {112}},\ \bibinfo {pages}
  {180404} (\bibinfo {year} {2014})}\BibitemShut {NoStop}%
\bibitem [{\citenamefont {Gerlach}\ and\ \citenamefont
  {Stern}(1922)}]{Gerlach22}%
  \BibitemOpen
  \bibfield  {author} {\bibinfo {author} {\bibfnamefont {W.}~\bibnamefont
  {Gerlach}}\ and\ \bibinfo {author} {\bibfnamefont {O.}~\bibnamefont
  {Stern}},\ }\bibfield  {title} {\enquote {\bibinfo {title} {Das magnetische
  moment des silberatoms},}\ }\href {\doibase 10.1007/BF01326984} {\bibfield
  {journal} {\bibinfo  {journal} {Z. Phys.}\ }\textbf {\bibinfo {volume} {9}},\
  \bibinfo {pages} {353--355} (\bibinfo {year} {1922})}\BibitemShut {NoStop}%
\bibitem [{\citenamefont {Wootters}\ and\ \citenamefont
  {Fields}(1989)}]{Wootters89}%
  \BibitemOpen
  \bibfield  {author} {\bibinfo {author} {\bibfnamefont {W.~K.}\ \bibnamefont
  {Wootters}}\ and\ \bibinfo {author} {\bibfnamefont {B.~D.}\ \bibnamefont
  {Fields}},\ }\bibfield  {title} {\enquote {\bibinfo {title} {Optimal
  state-determination by mutually unbiased measurements},}\ }\href {\doibase
  10.1016/0003-4916(89)90322-9} {\bibfield  {journal} {\bibinfo  {journal}
  {Ann. Phys.}\ }\textbf {\bibinfo {volume} {191}},\ \bibinfo {pages}
  {363--381} (\bibinfo {year} {1989})}\BibitemShut {NoStop}%
\bibitem [{\citenamefont {Paw\l{}owski}\ \emph {et~al.}(2009)\citenamefont
  {Paw\l{}owski}, \citenamefont {Paterek}, \citenamefont {Kaszlikowski},
  \citenamefont {Scarani}, \citenamefont {Winter},\ and\ \citenamefont
  {\.Zukowski}}]{Pawlowski2009}%
  \BibitemOpen
  \bibfield  {author} {\bibinfo {author} {\bibfnamefont {M.}~\bibnamefont
  {Paw\l{}owski}}, \bibinfo {author} {\bibfnamefont {T.}~\bibnamefont
  {Paterek}}, \bibinfo {author} {\bibfnamefont {D.}~\bibnamefont
  {Kaszlikowski}}, \bibinfo {author} {\bibfnamefont {V.}~\bibnamefont
  {Scarani}}, \bibinfo {author} {\bibfnamefont {A.}~\bibnamefont {Winter}}, \
  and\ \bibinfo {author} {\bibfnamefont {M.}~\bibnamefont {\.Zukowski}},\
  }\bibfield  {title} {\enquote {\bibinfo {title} {Information causality as a
  physical principle},}\ }\href {\doibase 10.1038/nature08400} {\bibfield
  {journal} {\bibinfo  {journal} {Nature (London)}\ }\textbf {\bibinfo {volume}
  {461}},\ \bibinfo {pages} {1101--1104} (\bibinfo {year} {2009})}\BibitemShut
  {NoStop}%
\bibitem [{\citenamefont {Oreshkov}\ \emph {et~al.}(2012)\citenamefont
  {Oreshkov}, \citenamefont {Costa},\ and\ \citenamefont
  {Brukner}}]{Oreshkov2012}%
  \BibitemOpen
  \bibfield  {author} {\bibinfo {author} {\bibfnamefont {O.}~\bibnamefont
  {Oreshkov}}, \bibinfo {author} {\bibfnamefont {F.}~\bibnamefont {Costa}}, \
  and\ \bibinfo {author} {\bibfnamefont {C.}~\bibnamefont {Brukner}},\
  }\bibfield  {title} {\enquote {\bibinfo {title} {Quantum correlations with no
  causal order},}\ }\href {\doibase 10.1038/ncomms2076} {\bibfield  {journal}
  {\bibinfo  {journal} {Nat. Commun.}\ }\textbf {\bibinfo {volume} {3}},\
  \bibinfo {pages} {1092} (\bibinfo {year} {2012})}\BibitemShut {NoStop}%
\bibitem [{\citenamefont {Brukner}(2014)}]{Brukner2014}%
  \BibitemOpen
  \bibfield  {author} {\bibinfo {author} {\bibfnamefont {C.}~\bibnamefont
  {Brukner}},\ }\bibfield  {title} {\enquote {\bibinfo {title} {Quantum
  causality},}\ }\href {http://dx.doi.org/10.1038/nphys2930} {\bibfield
  {journal} {\bibinfo  {journal} {Nat. Phys.}\ }\textbf {\bibinfo {volume}
  {10}},\ \bibinfo {pages} {259--263} (\bibinfo {year} {2014})},\ \bibinfo
  {note} {progress Article}\BibitemShut {NoStop}%
\bibitem [{\citenamefont {Procopio}\ \emph {et~al.}(2015)\citenamefont
  {Procopio}, \citenamefont {Moqanaki}, \citenamefont {Araujo}, \citenamefont
  {Costa}, \citenamefont {Alonso~Calafell}, \citenamefont {Dowd}, \citenamefont
  {Hamel}, \citenamefont {Rozema}, \citenamefont {Brukner},\ and\ \citenamefont
  {Walther}}]{Procopio2015}%
  \BibitemOpen
  \bibfield  {author} {\bibinfo {author} {\bibfnamefont {L.~M.}\ \bibnamefont
  {Procopio}}, \bibinfo {author} {\bibfnamefont {A.}~\bibnamefont {Moqanaki}},
  \bibinfo {author} {\bibfnamefont {M.}~\bibnamefont {Araujo}}, \bibinfo
  {author} {\bibfnamefont {F.}~\bibnamefont {Costa}}, \bibinfo {author}
  {\bibfnamefont {I.}~\bibnamefont {Alonso~Calafell}}, \bibinfo {author}
  {\bibfnamefont {E.~G.}\ \bibnamefont {Dowd}}, \bibinfo {author}
  {\bibfnamefont {D.~R.}\ \bibnamefont {Hamel}}, \bibinfo {author}
  {\bibfnamefont {L.~A.}\ \bibnamefont {Rozema}}, \bibinfo {author}
  {\bibfnamefont {C.}~\bibnamefont {Brukner}}, \ and\ \bibinfo {author}
  {\bibfnamefont {P.}~\bibnamefont {Walther}},\ }\bibfield  {title} {\enquote
  {\bibinfo {title} {Experimental superposition of orders of quantum gates},}\
  }\href {http://dx.doi.org/10.1038/ncomms8913} {\bibfield  {journal} {\bibinfo
   {journal} {Nat. Commun.}\ }\textbf {\bibinfo {volume} {6}} (\bibinfo {year}
  {2015})}\BibitemShut {NoStop}%
\bibitem [{\citenamefont {Ried}\ \emph {et~al.}(2015)\citenamefont {Ried},
  \citenamefont {Agnew}, \citenamefont {Vermeyden}, \citenamefont {Janzing},
  \citenamefont {Spekkens},\ and\ \citenamefont {Resch}}]{Ried2015}%
  \BibitemOpen
  \bibfield  {author} {\bibinfo {author} {\bibfnamefont {K.}~\bibnamefont
  {Ried}}, \bibinfo {author} {\bibfnamefont {M.}~\bibnamefont {Agnew}},
  \bibinfo {author} {\bibfnamefont {L.}~\bibnamefont {Vermeyden}}, \bibinfo
  {author} {\bibfnamefont {D.}~\bibnamefont {Janzing}}, \bibinfo {author}
  {\bibfnamefont {R.~W.}\ \bibnamefont {Spekkens}}, \ and\ \bibinfo {author}
  {\bibfnamefont {K.~J.}\ \bibnamefont {Resch}},\ }\bibfield  {title} {\enquote
  {\bibinfo {title} {A quantum advantage for inferring causal structure},}\
  }\href {http://dx.doi.org/10.1038/nphys3266} {\bibfield  {journal} {\bibinfo
  {journal} {Nat. Phys.}\ }\textbf {\bibinfo {volume} {11}},\ \bibinfo {pages}
  {414--420} (\bibinfo {year} {2015})},\ \bibinfo {note} {article}\BibitemShut
  {NoStop}%
\bibitem [{\citenamefont {Chaves}\ \emph {et~al.}(2015)\citenamefont {Chaves},
  \citenamefont {Majenz},\ and\ \citenamefont {Gross}}]{Chaves2015}%
  \BibitemOpen
  \bibfield  {author} {\bibinfo {author} {\bibfnamefont {R.}~\bibnamefont
  {Chaves}}, \bibinfo {author} {\bibfnamefont {Ch.}\ \bibnamefont {Majenz}}, \
  and\ \bibinfo {author} {\bibfnamefont {D.}~\bibnamefont {Gross}},\ }\bibfield
   {title} {\enquote {\bibinfo {title} {Information–-theoretic implications
  of quantum causal structures},}\ }\href {\doibase 10.1038/ncomms6766}
  {\bibfield  {journal} {\bibinfo  {journal} {Nat. Commun.}\ }\textbf {\bibinfo
  {volume} {6}},\ \bibinfo {pages} {5766} (\bibinfo {year} {2015})}\BibitemShut
  {NoStop}%
\bibitem [{\citenamefont {Gisin}\ \emph
  {et~al.}(2002{\natexlab{b}})\citenamefont {Gisin}, \citenamefont {Ribordy},
  \citenamefont {Tittel},\ and\ \citenamefont {Zbinden}}]{Gisin02}%
  \BibitemOpen
  \bibfield  {author} {\bibinfo {author} {\bibfnamefont {N.}~\bibnamefont
  {Gisin}}, \bibinfo {author} {\bibfnamefont {G.}~\bibnamefont {Ribordy}},
  \bibinfo {author} {\bibfnamefont {W.}~\bibnamefont {Tittel}}, \ and\ \bibinfo
  {author} {\bibfnamefont {H.}~\bibnamefont {Zbinden}},\ }\bibfield  {title}
  {\enquote {\bibinfo {title} {Quantum cryptography},}\ }\href {\doibase
  10.1103/RevModPhys.74.145} {\bibfield  {journal} {\bibinfo  {journal} {Rev.
  Mod. Phys.}\ }\textbf {\bibinfo {volume} {74}},\ \bibinfo {pages} {145--195}
  (\bibinfo {year} {2002}{\natexlab{b}})}\BibitemShut {NoStop}%
\bibitem [{\citenamefont {Soubusta}\ \emph {et~al.}(2007)\citenamefont
  {Soubusta}, \citenamefont {Bart\r{u}\v{s}kov\'{a}}, \citenamefont
  {\v{C}ernoch}, \citenamefont {Fiur\'{a}\v{s}ek},\ and\ \citenamefont
  {Du\v{s}ek}}]{Soubusta07}%
  \BibitemOpen
  \bibfield  {author} {\bibinfo {author} {\bibfnamefont {J.}~\bibnamefont
  {Soubusta}}, \bibinfo {author} {\bibfnamefont {L.}~\bibnamefont
  {Bart\r{u}\v{s}kov\'{a}}}, \bibinfo {author} {\bibfnamefont {A.}~\bibnamefont
  {\v{C}ernoch}}, \bibinfo {author} {\bibfnamefont {J.}~\bibnamefont
  {Fiur\'{a}\v{s}ek}}, \ and\ \bibinfo {author} {\bibfnamefont
  {M.}~\bibnamefont {Du\v{s}ek}},\ }\bibfield  {title} {\enquote {\bibinfo
  {title} {Several experimental realizations of symmetric phase-covariant
  quantum cloners of single-photon qubits},}\ }\href {\doibase
  10.1103/PhysRevA.76.042318} {\bibfield  {journal} {\bibinfo  {journal} {Phys.
  Rev. A}\ }\textbf {\bibinfo {volume} {76}},\ \bibinfo {pages} {042318}
  (\bibinfo {year} {2007})}\BibitemShut {NoStop}%
\bibitem [{\citenamefont {Bart\r{u}\v{s}kov\'{a}}\ \emph
  {et~al.}(2007)\citenamefont {Bart\r{u}\v{s}kov\'{a}}, \citenamefont
  {Du\v{s}ek}, \citenamefont {\v{C}ernoch}, \citenamefont {Soubusta},\ and\
  \citenamefont {Fiur\'{a}\v{s}ek}}]{Bartuskova07}%
  \BibitemOpen
  \bibfield  {author} {\bibinfo {author} {\bibfnamefont {L.}~\bibnamefont
  {Bart\r{u}\v{s}kov\'{a}}}, \bibinfo {author} {\bibfnamefont {M.}~\bibnamefont
  {Du\v{s}ek}}, \bibinfo {author} {\bibfnamefont {A.}~\bibnamefont
  {\v{C}ernoch}}, \bibinfo {author} {\bibfnamefont {J.}~\bibnamefont
  {Soubusta}}, \ and\ \bibinfo {author} {\bibfnamefont {J.}~\bibnamefont
  {Fiur\'{a}\v{s}ek}},\ }\bibfield  {title} {\enquote {\bibinfo {title}
  {Fiber-optics implementation of an asymmetric phase-covariant quantum
  cloner},}\ }\href {\doibase 10.1103/PhysRevLett.99.120505} {\bibfield
  {journal} {\bibinfo  {journal} {Phys. Rev. Lett.}\ }\textbf {\bibinfo
  {volume} {99}},\ \bibinfo {pages} {120505} (\bibinfo {year}
  {2007})}\BibitemShut {NoStop}%
\bibitem [{\citenamefont {Lemr}\ \emph {et~al.}(2012)\citenamefont {Lemr},
  \citenamefont {Bartkiewicz}, \citenamefont {\v{C}ernoch}, \citenamefont
  {Soubusta},\ and\ \citenamefont {Miranowicz}}]{Lemr12}%
  \BibitemOpen
  \bibfield  {author} {\bibinfo {author} {\bibfnamefont {K.}~\bibnamefont
  {Lemr}}, \bibinfo {author} {\bibfnamefont {K.}~\bibnamefont {Bartkiewicz}},
  \bibinfo {author} {\bibfnamefont {A.}~\bibnamefont {\v{C}ernoch}}, \bibinfo
  {author} {\bibfnamefont {J.}~\bibnamefont {Soubusta}}, \ and\ \bibinfo
  {author} {\bibfnamefont {A.}~\bibnamefont {Miranowicz}},\ }\bibfield  {title}
  {\enquote {\bibinfo {title} {Experimental linear-optical implementation of a
  multifunctional optimal qubit cloner},}\ }\href {\doibase
  10.1103/PhysRevA.85.050307} {\bibfield  {journal} {\bibinfo  {journal} {Phys.
  Rev. A}\ }\textbf {\bibinfo {volume} {85}},\ \bibinfo {pages} {050307}
  (\bibinfo {year} {2012})}\BibitemShut {NoStop}%
\bibitem [{\citenamefont {Bartkiewicz}\ \emph
  {et~al.}(2013{\natexlab{a}})\citenamefont {Bartkiewicz}, \citenamefont
  {Lemr}, \citenamefont {\v{C}ernoch}, \citenamefont {Soubusta},\ and\
  \citenamefont {Miranowicz}}]{Bartkiewicz13prl}%
  \BibitemOpen
  \bibfield  {author} {\bibinfo {author} {\bibfnamefont {K.}~\bibnamefont
  {Bartkiewicz}}, \bibinfo {author} {\bibfnamefont {K.}~\bibnamefont {Lemr}},
  \bibinfo {author} {\bibfnamefont {A.}~\bibnamefont {\v{C}ernoch}}, \bibinfo
  {author} {\bibfnamefont {J.}~\bibnamefont {Soubusta}}, \ and\ \bibinfo
  {author} {\bibfnamefont {A.}~\bibnamefont {Miranowicz}},\ }\bibfield  {title}
  {\enquote {\bibinfo {title} {Experimental eavesdropping based on optimal
  quantum cloning},}\ }\href {\doibase 10.1103/PhysRevLett.110.173601}
  {\bibfield  {journal} {\bibinfo  {journal} {Phys. Rev. Lett.}\ }\textbf
  {\bibinfo {volume} {110}},\ \bibinfo {pages} {173601} (\bibinfo {year}
  {2013}{\natexlab{a}})}\BibitemShut {NoStop}%
\bibitem [{\citenamefont {Wootters}\ and\ \citenamefont
  {Zurek}(1982)}]{Wootters82}%
  \BibitemOpen
  \bibfield  {author} {\bibinfo {author} {\bibfnamefont {W.~K.}\ \bibnamefont
  {Wootters}}\ and\ \bibinfo {author} {\bibfnamefont {W.~H.}\ \bibnamefont
  {Zurek}},\ }\bibfield  {title} {\enquote {\bibinfo {title} {A single quantum
  cannot be cloned},}\ }\href {\doibase 10.1038/299802a0} {\bibfield  {journal}
  {\bibinfo  {journal} {Nature (London)}\ }\textbf {\bibinfo {volume} {299}},\
  \bibinfo {pages} {802--803} (\bibinfo {year} {1982})}\BibitemShut {NoStop}%
\bibitem [{\citenamefont {Dieks}(1982)}]{Dieks82}%
  \BibitemOpen
  \bibfield  {author} {\bibinfo {author} {\bibfnamefont {D.G.B.J.}\
  \bibnamefont {Dieks}},\ }\bibfield  {title} {\enquote {\bibinfo {title}
  {Communication by {EPR} devices},}\ }\href {\doibase
  10.1016/0375-9601(82)90084-6} {\bibfield  {journal} {\bibinfo  {journal}
  {Phys. Lett. A}\ }\textbf {\bibinfo {volume} {92}},\ \bibinfo {pages}
  {271--272} (\bibinfo {year} {1982})}\BibitemShut {NoStop}%
\bibitem [{\citenamefont {Horst}\ \emph {et~al.}(2013)\citenamefont {Horst},
  \citenamefont {Bartkiewicz},\ and\ \citenamefont {Miranowicz}}]{Horst13}%
  \BibitemOpen
  \bibfield  {author} {\bibinfo {author} {\bibfnamefont {B.}~\bibnamefont
  {Horst}}, \bibinfo {author} {\bibfnamefont {K.}~\bibnamefont {Bartkiewicz}},
  \ and\ \bibinfo {author} {\bibfnamefont {A.}~\bibnamefont {Miranowicz}},\
  }\bibfield  {title} {\enquote {\bibinfo {title} {Two-qubit mixed states more
  entangled than pure states: Comparison of the relative entropy of
  entanglement for a given nonlocality},}\ }\href {\doibase
  10.1103/PhysRevA.87.042108} {\bibfield  {journal} {\bibinfo  {journal} {Phys.
  Rev. A}\ }\textbf {\bibinfo {volume} {87}},\ \bibinfo {pages} {042108}
  (\bibinfo {year} {2013})}\BibitemShut {NoStop}%
\bibitem [{\citenamefont {Bartkiewicz}\ \emph
  {et~al.}(2013{\natexlab{b}})\citenamefont {Bartkiewicz}, \citenamefont
  {Horst}, \citenamefont {Lemr},\ and\ \citenamefont
  {Miranowicz}}]{Bartkiewicz13PRA}%
  \BibitemOpen
  \bibfield  {author} {\bibinfo {author} {\bibfnamefont {K.}~\bibnamefont
  {Bartkiewicz}}, \bibinfo {author} {\bibfnamefont {B.}~\bibnamefont {Horst}},
  \bibinfo {author} {\bibfnamefont {K.}~\bibnamefont {Lemr}}, \ and\ \bibinfo
  {author} {\bibfnamefont {A.}~\bibnamefont {Miranowicz}},\ }\bibfield  {title}
  {\enquote {\bibinfo {title} {Entanglement estimation from {B}ell inequality
  violation},}\ }\href {\doibase 10.1103/PhysRevA.88.052105} {\bibfield
  {journal} {\bibinfo  {journal} {Phys. Rev. A}\ }\textbf {\bibinfo {volume}
  {88}},\ \bibinfo {pages} {052105} (\bibinfo {year}
  {2013}{\natexlab{b}})}\BibitemShut {NoStop}%
\bibitem [{\citenamefont {Chen}\ \emph {et~al.}(2013)\citenamefont {Chen},
  \citenamefont {Ye}, \citenamefont {Wu}, \citenamefont {Su}, \citenamefont
  {Cabello}, \citenamefont {Kwek},\ and\ \citenamefont {Oh}}]{Chen13}%
  \BibitemOpen
  \bibfield  {author} {\bibinfo {author} {\bibfnamefont {J.-L.}\ \bibnamefont
  {Chen}}, \bibinfo {author} {\bibfnamefont {X.-J.}\ \bibnamefont {Ye}},
  \bibinfo {author} {\bibfnamefont {C.}~\bibnamefont {Wu}}, \bibinfo {author}
  {\bibfnamefont {H.-Y.}\ \bibnamefont {Su}}, \bibinfo {author} {\bibfnamefont
  {A.}~\bibnamefont {Cabello}}, \bibinfo {author} {\bibfnamefont {L.~C.}\
  \bibnamefont {Kwek}}, \ and\ \bibinfo {author} {\bibfnamefont {C.~H.}\
  \bibnamefont {Oh}},\ }\bibfield  {title} {\enquote {\bibinfo {title}
  {All-versus-nothing proof of {E}instein-{P}odolsky-{R}osen steering},}\
  }\href {\doibase 10.1038/srep02143} {\bibfield  {journal} {\bibinfo
  {journal} {Sci. Rep.}\ }\textbf {\bibinfo {volume} {3}},\ \bibinfo {pages}
  {2143} (\bibinfo {year} {2013})}\BibitemShut {NoStop}%
\bibitem [{\citenamefont {Barnett}\ and\ \citenamefont
  {Dragomir}(1999)}]{Barnett99}%
  \BibitemOpen
  \bibfield  {author} {\bibinfo {author} {\bibfnamefont {N.~S.}\ \bibnamefont
  {Barnett}}\ and\ \bibinfo {author} {\bibfnamefont {S.~S.}\ \bibnamefont
  {Dragomir}},\ }\bibfield  {title} {\enquote {\bibinfo {title} {Some
  elementary inequalities for the expectation and variance of a random variable
  whose pdf is defined on a finite interval},}\ }\href@noop {} {\bibfield
  {journal} {\bibinfo  {journal} {RGMIA research report collection}\ }\textbf
  {\bibinfo {volume} {2}} (\bibinfo {year} {1999})}\BibitemShut {NoStop}%
\bibitem [{\citenamefont {Bhatia}\ and\ \citenamefont
  {Davis}(2000)}]{Bhatia00}%
  \BibitemOpen
  \bibfield  {author} {\bibinfo {author} {\bibfnamefont {R.}~\bibnamefont
  {Bhatia}}\ and\ \bibinfo {author} {\bibfnamefont {C.}~\bibnamefont {Davis}},\
  }\bibfield  {title} {\enquote {\bibinfo {title} {A better bound on the
  variance},}\ }\href {\doibase 10.2307/2589180} {\bibfield  {journal}
  {\bibinfo  {journal} {Am. Math. Monthly}\ }\textbf {\bibinfo {volume}
  {107}},\ \bibinfo {pages} {353--357} (\bibinfo {year} {2000})}\BibitemShut
  {NoStop}%
\bibitem [{\citenamefont {Grant}\ and\ \citenamefont {Boyd}(2008)}]{SDP1}%
  \BibitemOpen
  \bibfield  {author} {\bibinfo {author} {\bibfnamefont {M.}~\bibnamefont
  {Grant}}\ and\ \bibinfo {author} {\bibfnamefont {S.}~\bibnamefont {Boyd}},\
  }\href {http://cvxr.com/} {\enquote {\bibinfo {title} {{CVX}: {M}atlab
  software for disciplined convex programming},}\ } (\bibinfo {year}
  {2008})\BibitemShut {NoStop}%
\bibitem [{\citenamefont {Andersen}\ \emph {et~al.}(2014)\citenamefont
  {Andersen}, \citenamefont {Dahl},\ and\ \citenamefont {Vandenberghe}}]{SDP2}%
  \BibitemOpen
  \bibfield  {author} {\bibinfo {author} {\bibfnamefont {M.}~\bibnamefont
  {Andersen}}, \bibinfo {author} {\bibfnamefont {J.}~\bibnamefont {Dahl}}, \
  and\ \bibinfo {author} {\bibfnamefont {L.}~\bibnamefont {Vandenberghe}},\
  }\href {http://cvxopt.org/} {\enquote {\bibinfo {title} {{CVXOPT}: {P}ython
  software for convex optimization},}\ } (\bibinfo {year} {2014})\BibitemShut
  {NoStop}%
\bibitem [{\citenamefont {Sagnol}(2015)}]{SDP3}%
  \BibitemOpen
  \bibfield  {author} {\bibinfo {author} {\bibfnamefont {G.}~\bibnamefont
  {Sagnol}},\ }\href {http://picos.zib.de/} {\enquote {\bibinfo {title}
  {{PICOS}: A {P}ython interface for conic optimization solvers},}\ } (\bibinfo
  {year} {2015})\BibitemShut {NoStop}%
\bibitem [{\citenamefont {Bartkiewicz}\ \emph {et~al.}(2015)\citenamefont
  {Bartkiewicz}, \citenamefont {\v{C}ernoch}, \citenamefont {Lemr},
  \citenamefont {Miranowicz},\ and\ \citenamefont {Nori}}]{Karol15steering}%
  \BibitemOpen
  \bibfield  {author} {\bibinfo {author} {\bibfnamefont {K.}~\bibnamefont
  {Bartkiewicz}}, \bibinfo {author} {\bibfnamefont {A.}~\bibnamefont
  {\v{C}ernoch}}, \bibinfo {author} {\bibfnamefont {K.}~\bibnamefont {Lemr}},
  \bibinfo {author} {\bibfnamefont {A.}~\bibnamefont {Miranowicz}}, \ and\
  \bibinfo {author} {\bibfnamefont {F.}~\bibnamefont {Nori}},\ }\bibfield
  {title} {\enquote {\bibinfo {title} {Temporal steering and security of
  quantum key distribution with mutually-unbiased bases against individual
  attacks},}\ }\href@noop {} {\bibfield  {journal} {\bibinfo  {journal}
  {e-print arXiv:1503.00612}\ } (\bibinfo {year} {2015})}\BibitemShut {NoStop}%
\end{thebibliography}

%

\end{document}